\documentclass{llncs}
\pdfoutput=1
\usepackage{booktabs}

\usepackage{algorithm, algpseudocode}
\usepackage{relsize}

\renewcommand{\tabcolsep}{2pt}

\usepackage{epsfig, endnotes, url}
\usepackage{color, cite, graphicx}
\usepackage{mathtools, amsmath, amsfonts}
\usepackage{diagbox, booktabs, colortbl}
\usepackage{multirow, subcaption}
\usepackage{breakurl, listings, framed}
\usepackage[hidelinks]{hyperref}
\usepackage{xspace}
\usepackage{slashbox}
\usepackage{enumitem, array}
\DeclareMathAlphabet{\mathcal}{OMS}{cmsy}{m}{n}

\newcolumntype{?}{!{\vrule width 1pt}}

{
\makeatletter
\newcommand{\@chapapp}{\relax}%
\makeatother

\usepackage[title]{appendix}

\newcommand{\ignore}[1]{}

\usepackage{array}
\newcolumntype{P}[1]{>{\centering\arraybackslash}p{#1}}

\newcommand\sysname{CAPnet\xspace}

\hyphenation{op-tical net-works semi-conduc-tor}

\usepackage{ifthen}
\usepackage[normalem]{ulem} % for \sout
\usepackage{xcolor}
\usepackage{amssymb}

\newboolean{showedits}
\setboolean{showedits}{true} % toggle to show or hide edits
\ifthenelse{\boolean{showedits}}
{
	\newcommand{\del}[1]{\textcolor{red}{\sout{#1}}} % please delete
}{
	\newcommand{\del}[1]{} % please delete
	
}

\newboolean{showcomments}
\setboolean{showcomments}{true}
\newcommand{\id}[1]{$-$Id: scgPaper.tex 32478 2010-04-29 09:11:32Z oscar $-$}

\ifthenelse{\boolean{showcomments}}
{\newcommand{\nbc}[3]{
 {\colorbox{#3}{\bfseries\sffamily\scriptsize\textcolor{white}{#1}}}
 {\textcolor{#3}{\sf\small$\blacktriangleright$\textit{#2}$\blacktriangleleft$}}}
 }
{\newcommand{\nbc}[3]{}
 \renewcommand{\del}[1]{} % please delete
 }

\definecolor{jccolor}{rgb}{0.2,0.4,0.6}
\definecolor{clcolour}{rgb}{0.5,0.7,0.9}
\definecolor{kkcolor}{rgb}{0.6,0.4,0.2}

\usepackage{wasysym}

\begin{document}
\title{\Large \bf \sysname: A Defense Against Cache Accounting Attacks on Content Distribution Networks}

\author{Ghada Almashaqbeh\inst{1} \and
Kevin Kelley\inst{2} \and
Allison Bishop\inst{1,3} \and
Justin Cappos\inst{4}}
\authorrunning{G. Almashaqbeh et al.}
\institute{Columbia University, NY, USA
\email{\{ghada, allison\}@cs.columbia.edu}\\ 
\and 
CacheCash Development Company, CA, USA
\email{kelleyk@kelleyk.net}\\
\and
Proof Trading, NY, USA\\
\and
New York University, NY, USA
\email{jcappos@nyu.edu}}

\maketitle

% Use the following at camera-ready time to suppress page numbers.
% Comment it out when you first submit the paper for review.
%\thispagestyle{empty}

\begin{abstract}
Peer-assisted content distribution networks (CDNs) have emerged to 
improve performance and reduce deployment costs of traditional,
infrastructure-based content delivery networks. This is done by 
employing peer-to-peer data 
transfers to supplement the resources of the network infrastructure. 
However, these hybrid systems are vulnerable to accounting 
attacks in which the peers, or caches, collude with
clients in order to report that content was transferred when it was not. 
This is a particular issue in systems that
incentivize cache participation, because malicious caches may 
collect rewards from the content publishers
operating the CDN without doing any useful work.

In this paper, we introduce \sysname, the first technique that 
lets untrusted caches join a peer-assisted CDN
while providing a bound on the effectiveness of accounting attacks. 
At its heart is a lightweight \emph{cache accountability puzzle} that 
clients must solve before caches are given credit. This puzzle 
requires colocating the data a client has requested, so its solution 
confirms that the content (or at least an amount of data within 
a pre-configured bound) has actually been retrieved. We analyze
the security and overhead of our scheme in realistic scenarios. The 
results show that a modest client machine using a single core can solve
puzzles at a rate sufficient to simultaneously watch dozens of 1080p 
videos. The technique is designed to be even more
scalable on the server side. In our experiments, one core of a single 
low-end machine is able to generate puzzles for 4.26 Tbps of 
bandwidth --- enabling 870,000 clients to concurrently view the 
same 1080p video. This demonstrates that our scheme can ensure 
cache accountability without degrading system productivity.
\end{abstract}

\section{Introduction}
\label{intro}
Online content distribution has grown dramatically over the last 
decade. Video streaming, in particular accounts for more than 
60$\%$ of today's Internet traffic and it is 
expected to exceed 80\% by 2022 \cite{cisco-report}. To meet this 
huge demand, content publishers typically employ solutions that 
distribute the load among geographically dispersed caches. 
Among these solutions, infrastructure-based content delivery networks 
(CDNs) have proven effective\cite{akamai, cloudfront}. But as load 
continues to grow, pressure to improve performance and 
reduce cost has led to experimentation with more sophisticated 
topologies~\cite{Buyya2008, Jin2016}.

Peer-assisted CDNs have evolved to reduce these costs and also to 
allow access to lower latency peers.  In this model,
centralized services are supplemented with the resources of end 
users, or peers \cite{Zhao13, Anjum17}. By allowing
anyone to join, and exchanging service for a payment, this paradigm 
creates robust, performant, and flexible systems. In
addition, peer-assisted solutions can extend the network coverage 
of infrastructure-based CDNs, scale more easily with
demand, and when managed carefully, offer a good quality of service 
for end users \cite{Zhao13}.  Several commercial CDN
providers have built products that employ this approach, such as 
Swarmify \cite{swarmify}, Velocix \cite{velocix}, and
Akamai NetSession \cite{netsession}.

However, peer-assisted work models are vulnerable to cache accounting 
attacks~\cite{Lian07, Aditya12}, in which a cache
and client collude to defraud the content publisher by claiming to have 
transferred data (and claiming payment) when no
actual work has been done. This is a particular problem in content distribution 
applications that do not require subscription fees from clients, such as 
ad-funded video streaming~\cite{youtube}, or services that allow a client to 
play content on multiple devices under one subscription~\cite{netflix}.  
Some defenses against these attacks have been 
proposed~\cite{Aditya12, Reiter09}, but they
do not work in typical peer-to-peer scenarios, where untrusted, 
anonymous nodes serve as caches.

In this paper, we introduce \sysname, the first technique that 
lets untrusted caches, such as peers with unknown computational and latency 
characteristics, join a peer-assisted CDN while providing a bound on 
the effectiveness of accounting attacks. Our key innovation is a
lightweight \emph{cache accountability puzzle} that clients must solve 
before caches are given credit. The puzzle 
solution serves as a content retrieval confirmation assuring publishers 
that the claimed content transfer (or at least an amount of data within 
a pre-configured bound) has taken place.

For each service request in \sysname, the publisher generates a puzzle that a 
client must solve by processing the data chunks retrieved from caches (each 
of which is encrypted with a request-specific 
key). Solving this puzzle requires the client to sequentially touch small pieces of 
these chunks in an unpredictable order. Because of this unpredictability, 
the communication overhead of generating the solution without having the data
colocated is significant. Combined with the use of a \emph{completion mask}, a 
secret that is used to conceal an encrypted data
chunk until it has been completely transferred, this processing pattern forces 
colluding parties to expend bandwidth 
similar to retrieving the content, and thus removing any motivation to cheat.

Equally important for its use in practical applications, \sysname does not sacrifice 
efficiency for enhanced security. Its tools are built on computationally-light 
operations (symmetric encryption and 
hashing). \sysname is also designed to be scalable;
while a client needs to process a large portion of the retrieved content when 
solving a puzzle, a publisher needs only to process a small, server-configurable 
number of pieces to generate a challenge.  Because of this asymmetry, our scheme can
meet the deployment demands of large-scale content distribution applications.

To demonstrate that \sysname is effective at mitigating cache accounting 
attacks, we configure the system parameters
based on an analysis of the bandwidth cost incurred by malicious 
puzzle-solving strategies.  Our analysis shows that
the publisher can ensure that a malicious actor must expend a
substantial amount of bandwidth, even given unrealistically strong 
assumptions about the malicious actors capabilities.
To evaluate \sysname's efficiency, we experimentally evaluate the computational 
overhead of our scheme under
various configurations.  The benchmark results show that a modest client 
machine can solve puzzles at a rate
sufficient to confirm the retrieval of around 170 Mbps, which is enough to 
watch more than 30 1080p quality videos
simultaneously. Even a single core low-end publisher machine can 
generate enough puzzles to support 870,000 simultaneous views of 
the same video. 

\section{Related Work}
\label{related}
To orient readers to
current state-of-the-art defenses for cache accounting attacks,
this section reviews prior work done in this area. We also present
information about a related topic --- proofs of data storage --- 
and discuss why this proof paradigm is not applicable to 
cache accounting attacks. \\

\noindent{\bf Cache accountability in peer-assisted CDNs.}
One technique used in peer-assisted CDNs is to rely on
the peers themselves to report statistics about content delivery. 
For example, clients in Akamai Netsession~\cite{netsession}
share reports about their upload and download activity, and this
information is used to manage network resources. Even
some systems that exchange service for monetary rewards, 
e.g.,~\cite{Nair08}, 
rely on these types of reports to
track the service contributions of peers in order to pay
them accordingly.
However, in such an open environment
that allows anyone to join, peers may fabricate these
accounting reports. This has been confirmed through
empirical studies~\cite{Lian07, Aditya12}.

Consequently, specialized cache accountability defenses
work to address this issue by making clients commit to these activity logs.
This is done by requiring participants to maintain tamper-evident logs, and 
cryptographically sign all messages sent to the network. The participants
periodically exchange these logs with a verifier. The verifier in turn checks the 
consistency of the reported information 
and performs anomaly detection to identify cheating based on a protocol reference
implementation. The repeat and compare
scheme~\cite{Michalakis07} utilizes this technique to address the
problem of corrupted content distribution. PeerReview~\cite{Haeberlen07}
employs a similar approach to detect Byzantine faults. And
RCA (Reliable Client Accounting)~\cite{Aditya12} exploits such logs
to address the same collusion problem we describe in this paper.
However, this approach cannot prevent colluding parties
from fabricating consistent and valid-looking log reporting  
content transfers that did not take place. Thus, cheating clients and caches can
still collude to collect rewards for work they did not perform.

A prior bandwidth puzzle-based defense, proposed by Reiter et 
al.~\cite{Reiter09}, works by issuing challenge puzzles to all caches 
and clients (which must have known
computational abilities and communication latency) 
that possess the content.  These parties have to solve the issued puzzles
over the retrieved content in a short period of time to 
receive payment. This scheme has several downsides when compared to 
our approach. First, every time a new party retrieves the
content, puzzles must be solved by \emph{all} peers that have a copy of
this content, even those uninvolved in the transfer. Second, the
security of the scheme is based on knowing a bound for the
attacker's hashing power, which is used to quantify the number of challenge
puzzles that must be presented within a time window. Third, the latency
of peers must also be known for the scheme to resist cheating. This 
latency constraint may cause peers to lose their rewards in the event of 
lost or delayed messages. In addition, an attacker that can fool others into believing they
have high latency can cheat because she has more time to solve puzzles 
(including those that were supposed to be solved by other caches).  \\

\noindent{\bf Relation with data possession proofs.}
Several works in the literature have tackled a related problem
--- how to prove that a server to which a client has outsourced files is
actually storing those files, e.g., ensuring correct data storage in 
the cloud~\cite{Wang10}. Solutions to this problem
include proofs-of-retrievability~\cite{Cash17}, proofs of data 
possession~\cite{Erway15}, and proofs-of-storage~\cite{Dziembowski15}.  
Such proof systems, at first glance, could be viewed
as potential defenses against cache accounting attacks. That is, a 
publisher can ask a client to prove the storage of a local copy of the
retrieved content.
However, this does not confirm that caches have served the content.
These colluding caches can generate valid proofs of storage for any client
because they store the full raw content. Similarly, a client that retrieves
some content only once can produce valid proofs, for itself or others, for
all future requests that ask for the same content. While useful, these proof
systems are not applicable for fighting cache accounting attacks.

\section{\sysname Design}
\label{design}
\sysname defends against cache accounting 
attacks by both mandating proof of delivery, and making honest 
choices more profitable than cheating. In this section, we describe 
how the design of \sysname manages this defense. We start by 
defining the work environment for content distribution systems 
our scheme targets, then provide a high level 
view of the primary operations, after which we present  
the technical details of these operations.

\subsection{Work Environment Model}
In the incentivized, peer-assisted content distribution 
systems that \sysname targets, there are three participant
types: publishers, caches, and clients. A \emph{publisher} owns 
content, e.g., videos or software packages, that \emph{clients}
want to retrieve. A publisher hires \emph{caches} to distribute this 
content in exchange for rewards, such as monetary
incentives, which are tied to the amount of service these caches provide.  
Each cache is defined by its IP address, which the publisher 
monitors to detect Sybils. When a cache joins a publisher's network, it
gains access to the content to be served, which we assume to be divided 
into equally-sized data chunks. A client request can fetch a range of $n$  
chunks within the object, e.g., movie, it wants to retrieve.

During the content distribution process, a publisher acts as a
dispatcher assigning caches to fulfill client requests. 
Therefore, clients must contact the publisher first, asking for 
$n$ data chunks, to obtain 
the list of $n$ caches that will provide the service. The 
publisher selects this set randomly such 
that each cache will serve a single data chunk among the requested 
set.

As will be shown shortly, confirming that the content has been 
retrieved is done over individual content requests. In other words, even 
if the client wants to retrieve a large object, e.g., a movie of size 1 GB, it computes 
the retrieval 
confirmation over each $n$ chunks separately. Such an approach reduces the 
amount of memory that \sysname requires, e.g., for $n = 4$ and a 
chunk size of 1 MB, a client/publisher would need only a 4 MB storage to hold the chunks 
needed for processing any request.

\sysname enables the
publisher to set a bound on the amount of bandwidth the attacker
must expend, with respect to the 
original content amount, which we call the \emph{$\delta$-bound}. So, for 4MB 
of content, a $0.75$-bound attacker in our scheme 
is expected to expend 3MB to provide a valid content retrieval confirmation for a 
content request. The $\delta$-bound is controlled by the number of 
rounds in the cache accountability puzzle of \sysname. The larger $\delta$, the 
larger the computation cost of generating and solving this puzzle. Therefore, system 
designers need to configure this parameter based on the desired 
security-efficiency trade-off they want to achieve.

Lastly, we work in the random oracle model, where hash functions are modeled 
as random oracles. We also work in the ideal cipher model, where a block cipher 
is modeled as a random permutation. In addition, we deal with 
efficient adversaries that cannot break secure cryptographic primitives, 
such as AES, SHA256, and pseudorandom functions (PRFs), with 
non-negligible probability.

\subsection{Overview}
\sysname consists of a set of actions integrated into the content
delivery process. Collectively, these actions demonstrate that 
the required bandwidth amount  
was actually expended, even in the face of malicious,
colluding client and caches. In what follows, we provide an intuitive 
discussion of these actions to highlight the motivation behind 
the design (a rigorous security analysis of how these actions 
defend against cache accounting attacks is found in the next section).

\begin{figure}[t!]
\centerline{
\includegraphics[height= 2.1in, width =0.85\columnwidth]{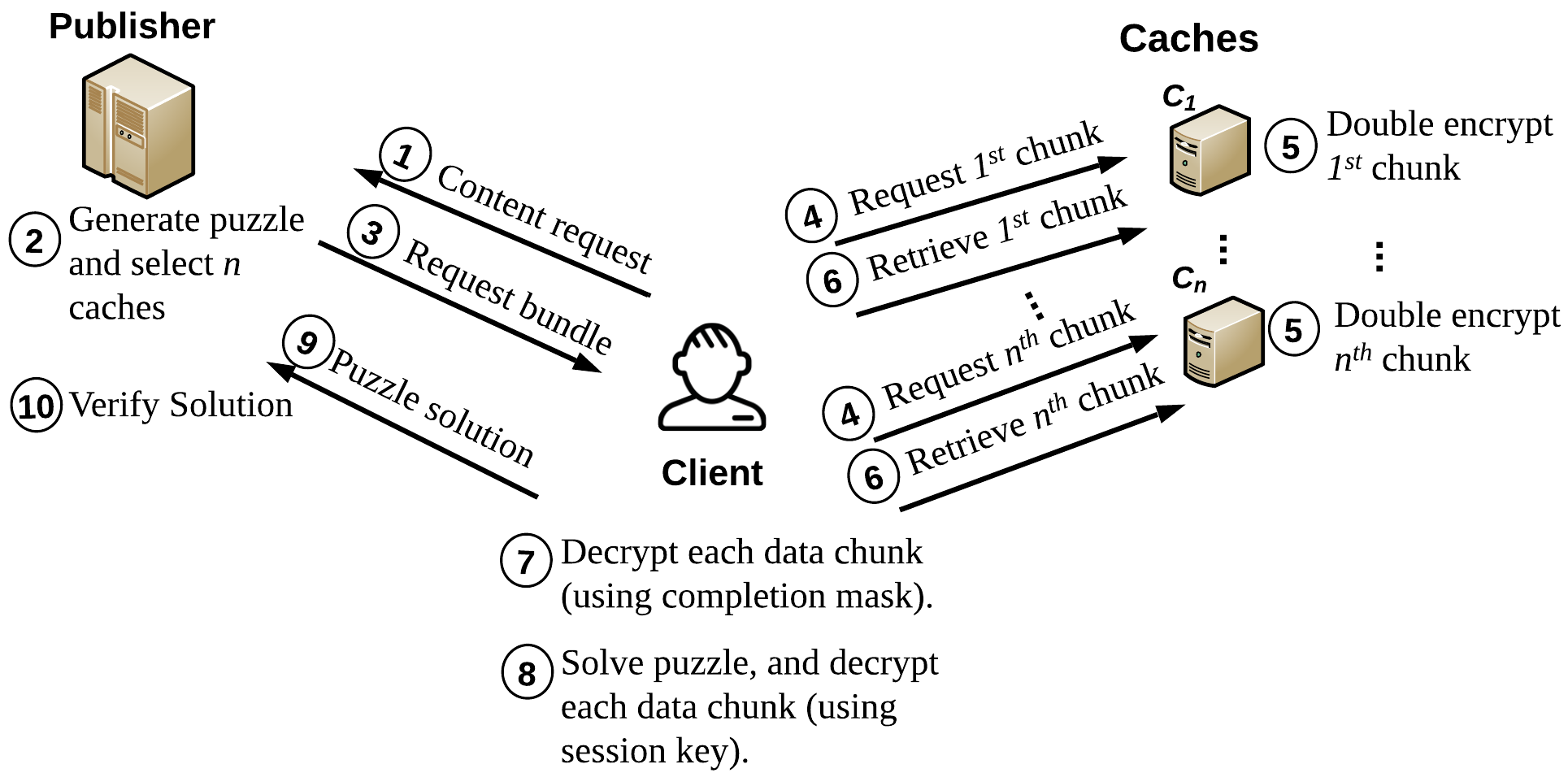}}
\caption{\sysname integration in content distribution ($n$ is the number of caches selected for a service request).}
\label{service-session}
\vspace{-6pt}
\end{figure}

As shown in Figure~\ref{service-session}, to request content 
in the \sysname model, a client retrieves a request bundle from the 
publisher that enables the retrieval of $n$ data chunks 
({\bf steps 1, 2,} and {\bf 3})\footnote{If the 
content has more than $n$ chunks a client will send several 
requests.}. This bundle stipulates 
which caches to contact, and includes the client's IP and a request number. In 
addition, the bundle
contains a puzzle, that when solved, enables the client to prove that
the requested chunks (or at least an amount of data within the
$\delta$-bound) were indeed retrieved.

The client contacts caches, possibly in parallel, and provides the request 
bundle that instructs each
cache what specific data chunk to serve ({\bf step 4}). Each cache will 
encrypt its data chunk with a unique per-request key, and additionally
encrypts the produced ciphertext using 
a fresh per-request completion mask ({\bf step 5}). The double 
encrypted chunk, appended with the
completion mask, is then delivered to the client ({\bf step 6}). Once all 
chunks are received, the client decrypts each chunk using the 
completion mask as the decryption key ({\bf step 7}), and solves the 
puzzle using the single-layer encrypted chunks   
({\bf step 8}). With the puzzle solution the
client can decrypt the data chunks to obtain the raw content ({\bf step 8}), and
confirm to the publisher that these chunks were retrieved
({\bf steps 9} and {\bf 10}).

When a cache begins serving content for a publisher, they 
establish a shared secret called a master key. Along with the
request number, both parties use this key to non-interactively 
generate a fresh per-request session key that is used to encrypt the
data chunk the cache will serve. Since this key is unique and cannot 
be distinguished from random\footnote{This is by the security of 
PRFs.}, each 
request returns a different, random looking, encrypted 
chunk\footnote{Recall that we 
work in the ideal cipher model.}, even if the raw 
content is the same.  Also, since the session key is a secret 
known only to the publisher and that cache, a malicious
cache does not know the encrypted content that an honest cache 
would serve.

In addition, \sysname ensures that a client retrieves
the entire encrypted chunk before it can start solving the puzzle. To do this, 
for each request a cache selects a random \emph{completion mask}, 
i.e., a random key, that is used to encrypt the chunk ciphertext. 
In other words, the cache adds a second layer of encryption using the 
completion mask as the encryption key. The 
cache appends the completion mask to the transmitted, double 
encrypted chunk. Thus, the 
client has to download the entire chunk before being 
able to unmask any part of it.

A puzzle is computed by processing small 
(e.g., 16 byte long) \emph{pieces} of the (single-layer) encrypted 
chunks. Starting at a 
randomly selected piece in the first chunk, one computes the  
hash of this piece and maps it to  
a piece index in the second data chunk. In the random oracle model, 
this mapping randomly
jumps to a piece in the second chunk. The hash is now
computed over the previous hash and the second piece and is used to
select another piece in the third chunk, and so on. Once the data chunk 
from each cache is visited, this completes a \emph{round}. The next 
round is begun by mapping the last hash of the prior round to a piece index 
in the first chunk. This continues for the number of 
rounds chosen by the publisher to achieve the desired $\delta$-bound.

The publisher and the client compute the puzzle in slightly different ways.
The publisher randomly chooses a ``starting piece'' in the
first chunk and computes one puzzle to produce a challenge for the client. This
challenge does not contain any information about the starting piece. Hence, 
the client will attempt
to solve the challenge by computing candidate, or trial, puzzles initiated at various
starting pieces in the first chunk until the solution is found. This forces the client 
to process a $\delta$ portion of the content before finding the solution.

Increasing $\delta$ strengthens the security 
guarantees of \sysname by causing malicious parties to retrieve more content, 
but also increases the computation cost for publishers and honest clients as 
larger number of rounds are needed. Caches, on the other hand, have uniform
computational cost independent of the $\delta$-bound.

\subsection{Design Details}
We now describe the \sysname actions in more detail,
including puzzle generation, solving, and verification.

\subsubsection{Puzzle Generation}
\begin{figure}[t!]
\centerline{
\includegraphics[height= 2.1in, width = 0.75\columnwidth]{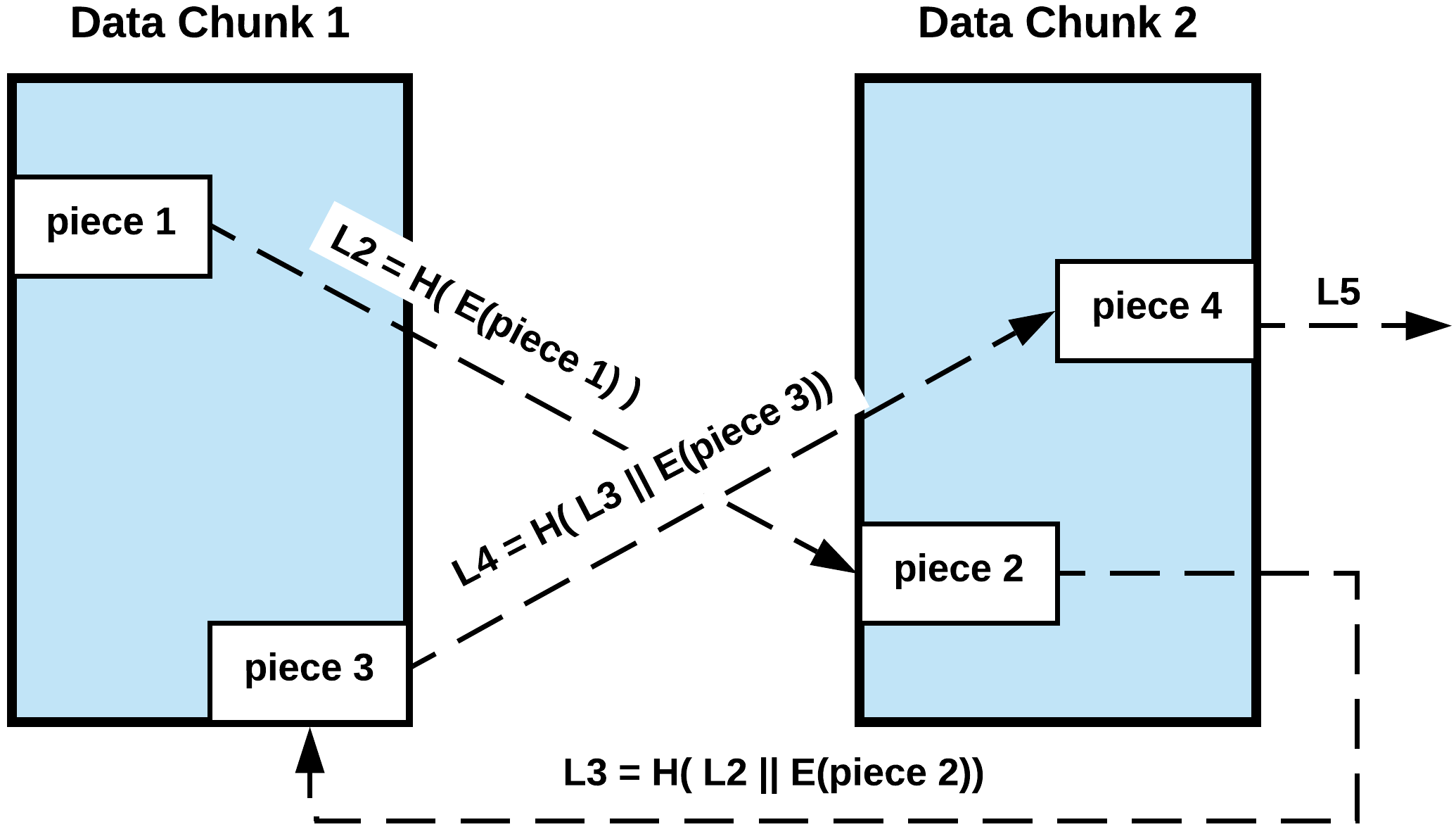}}
\caption{An example of puzzle challenge generation with two chunks and two
rounds. The puzzle challenge is $H(L_5)$ and its solution is $L_5$.}
\label{puzzle-diagram}
\end{figure}

The publisher generates a
challenge puzzle based on the data chunks a client wants to
retrieve. Figure~\ref{puzzle-diagram} depicts this action through
a clarifying example involving two data chunks. In this figure,
$L$ stands for location, $H$ stands
for hashing, $E$ stands for encryption, $||$ is a
concatenation operation, and the arrows indicate the sequence
of pieces selected when computing a puzzle.

As shown, a puzzle starts at the first
data chunk and proceeds by processing a number of small data
pieces selected at random. Given that a 
puzzle round processes encrypted pieces, the publisher 
encrypts the piece selected at each step. It then 
hashes the encrypted piece along with prior hash or location value. The
output hash is mapped to a piece index within the next chunk (or 
the first chunk is this is the beginning of a new round).

This computation pattern imposes three aspects. First, each 
location value encapsulates all encrypted pieces
processed so far, which enforces sequential computation of the
puzzle. Second, processing encrypted pieces
prevents any correlation between puzzles generated
for different requests, even if they are for the same raw 
content. Third, touching all data chunks in a round robin order  
makes all chunks contribute equally in solving a puzzle. This ensures  
that a puzzle solution is computed over all chunks, confirming  
their retrieval.

Once the puzzle computation is completed, which happens 
when the designated number of rounds is reached, the publisher uses the 
hash of the last location, i.e., $H(L_5)$ in the
figure, as the challenge, and it asks the client
to return the preimage of this hash, i.e., $L_5$. This is done
without revealing $piece_1$. Instead, the client tries all data pieces
in the first chunk. Hiding the starting piece causes the client to touch 
a large percentage of the pieces in each chunk 
while enabling the publisher to touch very few. This minimizes the
computational load for publishers, allowing them to process a
large number of client requests concurrently.

The technical details of this process are captured by
Algorithm~\ref{concrete-puzzle-origin}. In this algorithm,
$D_j$ is the $j^{th}$ data chunk, $piece_i$ is the $i^{th}$
data piece in a data chunk, $R_{puzzle}$
is the number of puzzle rounds, and $pieces_{total}$ is
the total number of pieces in any data chunk. The data
pieces inside a chunk are referenced using their indices, where the first piece 
has an index 0, the second has an index 1, and so on. We use $index(\cdot)$
to denote the index of a given piece.

\begin{algorithm}[t!]
\caption{Puzzle challenge generator.} \label{concrete-puzzle-origin}
\begin{algorithmic}[1]
\small
\State {\textbf{Input:} Data chunks $D_1, \dots, D_n$}

\State {\textbf{Output:} Challenge}

\\\Comment{ Initialization }
\For{$j$ = 1 to $n$}
\State {Generate $k_j$ and $ctr_{j, initial}$}
\EndFor
\\\Comment{ Puzzle generation }
\State{Select $index(piece_1)$ randomly from $D_1$}
\State{Set $j=1$, $L_1 = 0$, $r = n \cdot R_{puzzle}$}
\For{$i$ = 1 to $r$}
\State{Fetch $piece_i$ from $D_j$ based on $index(piece_i)$}
\\$\;\;\;\;$\Comment{ Compute $ctr_i$ and encrypt $piece_i$ }
\State{$ctr_i = index(piece_i) + ctr_{j, initial}$}
\State{$c_i = \mbox{AES-CTR}(k_j, ctr_i, piece_i)$}
\\$\;\;\;\;$\Comment{ Compute location and index of $piece_{i+1}$ }
\State{$L_{i+1} = \mbox{SHA}256(L_i||c_i)$}
\State{$index(piece_{i+1}) = L_{i+1} \mod pieces_{total}$}
\State{$j = (j \mod n) + 1$}
\EndFor

\State{Set Challenge $ = \mbox{SHA}256(L_{r+1})$}

\State{{\bf return} Challenge}
\end{algorithmic}
\end{algorithm}

Algorithm~\ref{concrete-puzzle-origin} shows two phases:
an initialization phase, and a challenge preparation phase.
The initialization phase is needed to allow caches and the publisher
to agree on the encryption setup. We use AES
in the counter mode (AES-CTR) for encryption because it allows
a publisher to encrypt any individual data piece without encrypting
the whole data chunk. To generate identical encrypted data,
the publisher and each cache $C_j$ must
generate the same session key $k_j$ and the initial value of
the AES-CTR counter $ctr_{j, initial}$.

Generating the AES-CTR counter and session key is done 
using a one time setup  
without any per-request interaction between the publisher and cache.
As mentioned before, a cache $C_j$ shares a master key with the 
publisher that they both use to derive
any future session key $k_j$. This is done by means of a pseudorandom 
function (PRF) keyed with this master key, and evaluated over the request 
number and the client IP to output $k_j$. The same PRF idea, but 
keyed with the session key, is used to generate $ctr_{j, initial}$.
Accordingly, in the initialization phase, the publisher generates
all keys and counters for all data chunks that will be used by the caches 
involved in the service session.

The puzzle generation phase proceeds as described previously. 
After selecting a piece index at random from
the first data chunk, the publisher proceeds by
computing the location of the next piece (line 16), and
mapping this location to a piece index (line 17). The new location computation
requires encrypting the prior piece, which in turn
requires computing the correct AES-CTR counter value (line 13).
By doing so, the publisher produces the same ciphertext of the 
selected piece that a cache 
will produce. The aforementioned process is repeated
for the required number of iterations, as found in lines 10-19. Lastly,
the algorithm outputs the hash of the last location as the puzzle
challenge.

After the puzzle is generated, the publisher informs the client
about the puzzle challenge it has to solve as part of the
request bundle mentioned earlier. To allow the client to
decrypt the retrieved data chunks, the publisher can either provide the keys in 
response to the puzzle solution reported by the client, or simply encrypt all 
session keys using the puzzle solution and share the ciphertext 
as part of the request bundle. Either way, once the client solves the challenge
puzzle it can recover these keys and decrypt the
received data.

\subsubsection{Puzzle Solving}
The client receives the puzzle challenge, along with the cache contact 
information, within the request bundle sent by the publisher. With this bundle, 
the client can start the content retrieval process, where it connects with 
the listed caches and requests the specified data chunks. Caches  
will deliver double-layer encrypted chunks, with the 
completion mask appended 
to each chunk. The client uses the completion mask as the decryption key 
to remove the second encryption layer of the chunk. By repeating this process 
for all chunks, the client obtains the 
single-layer encrypted data chunks.

The client can now perform the
second action in the \sysname process --- solving
the challenge puzzle. It uses a similar algorithm to the one used by 
the publisher with
three differences. First, since the client retrieves encrypted data
chunks from the caches, it does not encrypt the data
pieces before applying the hash, thus skipping lines 4-6 and 13-14 
in Algorithm~\ref{concrete-puzzle-origin}.
Second, since the client does not know the starting piece, it computes a
puzzle for every piece in the first data chunk until the correct solution is
found. In other words, it repeats lines 9-11 for each candidate 
starting piece. And third, once the client solves the puzzle, the
output is the puzzle solution, which is the last location
in the correct puzzle.

\vspace{-6pt}
\subsubsection{Puzzle Verification}
The last action in the \sysname process is verifying the
correctness of the reported puzzle solution.  While it would be possible
to keep a record of the puzzle challenges and their solutions for 
each client, we devise a 
computationally-lightweight technique that does not require maintaining
any per-client state.

In this technique, the publisher  
generates a unique secret token for each content request. This is done   
by evaluating a secret PRF over
the concatenation of the request number and the client IP. The publisher then 
encrypts the secret token using the puzzle solution,
and sends the encrypted token to the
client as part of the request bundle. Once
the client solves the puzzle, it can decrypt the token and send it back to the publisher
along with the request number. The publisher can simply evaluate
the secret PRF over the request number and the client IP, and 
thus, verifies whether the
output equals to the token value reported by the client. This enables the publisher 
to quickly check the correctness of a puzzle solution.

\section{Security Analysis}
\label{security-analysis}
\vspace{-8pt}
In this section, we analyze the effectiveness of \sysname
in fighting cache accounting attacks.  We begin by
outlining the setup of this analysis, after which we discuss how 
the security of \sysname changes as the design parameters change. 

\vspace{-6pt}
\subsection{Setup}
\vspace{-6pt}
The analysis setup defines how we model our adversaries, and explains 
the security properties that \sysname is designed to achieve.
The set of notations that this analysis uses is shown in 
Table~\ref{notation}.\\

\noindent{\it \underline{Adversary Model.}}
We consider a client colluding with a set of $m \geq 1$ caches\footnote{All 
these caches are different, i.e., not Sybils run on the same machine. This is 
due to the assumption that a publisher monitors the IP addresses of its caches 
to detect Sybils.}. (If a client
does not collude with any cache, it must retrieve all the data chunks 
to solve the puzzle just like an honest client.)  This
collusion can be modeled as an interaction between two parties: the
client and a collective entity $\mathbf{C}_m$. Any cache $C_j$ in
$\mathbf{C}_m$ can pool
all encrypted data chunks from the rest of the malicious caches
at a very low cost. That is, given that each cache has a full copy of the
raw content, $C_j$ needs only the session keys of these caches
to produce their encrypted data chunks locally. When we say
that a client retrieves data pieces from $\mathbf{C}_m$, we mean that this
client is interacting with the cache that pooled the data chunks.

\begin{table}[t!]
\caption{Notations.} 
\label{notation}
\centering \small{
\begin{tabular}{| p{0.16\columnwidth} | p{0.81\columnwidth} |}\hline\hline
{\bf Symbol} & {\bf Meaning}  \\ [0.5ex] \hline\hline

$n$ & Total number of caches in a service session.  \\[0.5ex] \hline
$\mathbf{C}_m$ & Set of malicious caches in a service session. \\ [0.5ex]  \hline  
$m$ & Number of malicious caches in a service session, where $m \leq n$.    \\ [0.5ex]  \hline  
$h_{size}$ & Hash output size. \\ [0.5ex]  \hline 
$chunk_{size}$ & Data chunk size. \\ [0.5ex]  \hline
$piece_{size}$ & Data piece size, where $piece_{size} \leq \frac{h_{size}}{m}$. \\ [0.5ex]  \hline
$pieces_{total}$ & Total number of pieces in a data chunk, where 
$pieces_{total} = \frac{chunk_{size}}{piece_{size}}$. \\ [1ex]  \hline
$R_{puzzle}$ & Number of puzzle rounds. \\ [0.5ex]  \hline
$Y$ & A random variable that represents the number of pieces
a malicious puzzle solver retrieves. \\ [0.5ex]  \hline
$\mathbb{E}[Y]$ & The expectation of $Y$. \\ [0.5ex]  \hline
$\delta$ & The ratio between the bandwidth amount a malicious 
puzzle solver would spend and
the amount that an honest solver would use. This is computed as 
$\delta = \frac{\mathbb{E}[Y]}{n \cdot pieces_{total}}$. \\ [1ex]  \hline

\end{tabular}}
\end{table}

In order to have a strong bound on attacker capabilities, we consider an 
attacker with full knowledge about the piece distribution across 
all the trial puzzles a client will compute. In other words, the attacker knows the 
selection frequency of data pieces, i.e., how many times a piece has 
been processed by all puzzles, in all
chunks rather than just in the puzzles for which this attacker has enough 
prior pieces. The attacker may use this information to retrieve the most 
frequent pieces when solving the challenge puzzle.

Despite the hash
function being modeled as a random oracle with a uniform and random 
output, this piece frequency still matters.  Suppose that
we have a chunk composed of 4 pieces that are randomly chosen to be
in 4 trial puzzles.  Over 90\% of the time one of these pieces is chosen
at least twice (only about 9\% of random draws of 4 items choose one from
each). On average more than 1 piece is likely not to be chosen for any
trial puzzle.  An intelligent attacker would choose the piece used in the most trial puzzles
since it gives the greatest chance to solve the challenge. So, accounting
for the fact that the actual frequencies are not uniform, even with a random function, more
accurately models the attacker's capabilities. Furthermore, providing the attacker perfect
information about these actual frequencies implies that the security
bound we infer will be conservative.

The client and $\mathbf{C}_m$ want to solve the puzzle while expending as little
bandwidth as possible. In quantifying this cost, we compute
the download bandwidth consumption for the colluding group.

Our adversary model is subject to the following assumptions:
\begin{enumerate}[label=A\arabic*.]
\itemsep0em
\vspace{-2pt}
\item {\bf Secure cryptographic primitives.} Adversaries cannot efficiently 
break the basic cryptographic building blocks (SHA256, AES, and PRFs) 
with non-negligible probability.

\item {\bf Clients do not already possess the content.} At the
beginning of a service session, a client does not have a copy of the
content it will request. This can be achieved by
having publishers track which clients have retrieved
which content. However, even if this assumption is violated,
the client still must retrieve data chunks from honest caches in
order to solve a puzzle, leading to the retrieval of at least
$\delta = \frac{n-m}{n}$ of the requested chunks.

\item {\bf Free adversarial metadata communication.} It is difficult to know
the minimal size of information adversaries would need to communicate when
coordinating the puzzle solving process.  Therefore, we will just assume
that such costs are free from a bandwidth standpoint and only count data piece
transmission. While it ignores some cost, this makes the overhead numbers
conservative in that real attackers will incur \emph{more cost} than what we
predict.

\item {\bf Content is already compressed.} The raw content distributed by caches 
is already compressed. As such, a malicious cache who intends to 
compress the raw content before encrypting it will only save a very small 
bandwidth amount.
\end{enumerate}

In addition to the above, and as mentioned previously, we work in 
the random oracle model (i.e., hash functions are modeled as random oracles), and 
in the ideal cipher model (i.e., block ciphers are modeled as random 
permutations).

An intelligent client and $\mathbf{C}_m$ collaborate to solve
the puzzle while transferring the least amount of data
possible. In this collaboration, the client receives encrypted chunks 
only from honest caches, while $\mathbf{C}_m$ produce all encrypted
chunks of malicious caches by pooling their session keys as
explained previously. (We consider that $\mathbf{C}_m$ have pooled 
the data to simplify our analysis.)  The strategy
then will have a \emph{solver}, either the client or $\mathbf{C}_m$, that
attempts to solve the puzzle using the
chunks it has in addition to information it requests from the second
party, whichever of the client or $\mathbf{C}_m$ is not the solver.
The second party acts as a \emph{piece provider} which sends 
pieces or hashes (i.e., piece locations computed in a puzzle round) 
to the solver. For reasons we will see shortly, pieces make more sense
for the attacker to transmit.

As our analysis will show, the client and $\mathbf{C}_m$ can decide in advance 
which party will play which role based on which option incurs the least
bandwidth cost.  This decision depends on the number of malicious
caches $m$.  If this number is less than half, i.e., $m < \frac{n}{2}$, 
it is more
efficient for the client to act as the solver.  If it is greater than half,
it is more efficient for the client to let $\mathbf{C}_m$ solve the 
challenge puzzle.  If the number of caches is exactly half, it is equally efficient
regardless of who is the solver.

Other attack strategies may involve attacking the cryptographic 
primitives used in CAPnet design. This may include trying to find 
the preimage of the puzzle challenge by inverting the hash, or predicting the 
session keys used by honest caches to produce their encrypted chunks 
locally by a colluding cache. By the 
security of the underlying cryptographic primitives CAPnet employs, 
i.e., the use of secure PRFs and first-preimage resistant hash functions, 
such strategies will succeed with negligible probability.

It should be noted that the above list is not known to be 
comprehensive. There could be other attack strategies outside the 
analysis presented in this section (although we are not aware of 
any of such strategies). Performing a rigorous analysis over the full 
potential threat space is an open problem for future work.\\

\noindent{\it \underline{Security Goal.}}
The goal is to ensure that a malicious puzzle solver cannot solve the challenge
puzzle in \sysname unless it expends, on average, a bandwidth amount 
equivalent to retrieving at least $\delta$ portion of the content.
This means that the colluding group is expected to 
expend a total of $n\delta chunk_{size}$ bandwidth units.  So for $\delta=0.95$, 
the attacker has an expected value of 95\% of the bandwidth cost even if all 
metadata overhead are ignored.  System designers
may set the value of $\delta$
based on the security-efficiency trade-off they want to achieve.
A larger $\delta$ value provides stronger security guarantees,
but also increases the computational cost of generating and solving puzzles.

It is not practical to have $\delta = 1$
unless the publisher touches every piece of the requested chunks.  Since
each chunk must be encrypted with a fresh key, this 
cost is prohibitive. In fact, if the publisher is willing to touch (and encrypt) every piece, 
it is simpler to compute the hash of the encrypted chunks, and use this 
hash as the content retrieval confirmation 
that a client has to compute. However,
this would greatly reduce performance. Assuming that the publisher does not
touch every piece, then $\delta < 1$ for the following reason.  Suppose that
the attacker retrieves
every piece of the content except one. If this piece was not touched 
by the publisher, the attacker can prove that the content was retrieved with 
$\delta < 1$.  Since we only account
for the piece transfer costs, at least some of the time (when the attacker does not
retrieve untouched pieces) $\delta < 1$, which makes the expected value 
of all cases to be less than 1.

\subsection{Analysis of Puzzle Solving Strategies}
In what follows, we analyze the bandwidth
cost of the malicious puzzle solving strategies described 
earlier, and show how to configure 
\sysname's design parameters to achieve the desired 
$\delta$-bound. These parameters include the data piece size 
$piece_{size}$ and the number of puzzle rounds $R_{puzzle}$.

As mentioned previously, an attacker who wishes to solve 
the puzzle without retrieving all the data chunks will 
either exchange hashes or retrieve data pieces. By
setting the $piece_{size} \leq \frac{h_{size}}{m}$ one can ensure that the 
cost of transmitting a hash is no less than transmitting pieces. That is, 
even in the event when the piece provider has $m$ consecutive encrypted 
chunks, meaning that given one hash value the provider can process $m$ pieces 
in a puzzle round, transmitting a hash is more expensive than transmitting 
these $m$ pieces. In fact since the pieces may be used in multiple puzzle 
trials, it is better for the solver to retrieve them. For this reason, we focus 
on strategies that involve piece dissemination instead.

When retrieving data pieces to solve the challenge puzzle, we 
conjecture that the 
best strategy for the solver is to utilize its knowledge of the 
piece distribution across all trial puzzles. Initially, the solver  
must possess some piece of each encrypted data chunk to have a chance to solve 
the puzzle, since each round touches all encrypted data chunks.  In
order to get pieces from the honest caches, the 
client must download all double-layer encrypted chunks held by 
these caches. For malicious caches, the client can retrieve the 
individual pieces it desires. In selecting which pieces to retrieve, the best 
approach is to ask for the piece that gives the greatest chance of solving 
the puzzle.  The solver can ask the piece provider to send the 
piece with the highest frequency among the 
remaining pieces, and then determine if it enables solving the 
challenge puzzle. This process 
continues until the solution is found. 
Assuming that retrieving the most popular missing piece 
is optimal, this is the optimal strategy
for the malicious solver.

Recall that either the client or a cache ($\mathbf{C}_m$)
may play the role of the puzzle solver.  If a cache is the puzzle solver, 
the client must still be the one to download the chunks from honest caches
since the source IP of the request is checked.  Hence, the client retrieval
from honest caches is a fixed cost.  Once this happens, it
is more efficient for the party with the most content (either the client or 
$\mathbf{C}_m$) to act as the solver and get as few pieces as possible from the other 
party. This means that when $m < \frac{n}{2}$, 
the client will have a larger number of chunks than $\mathbf{C}_m$, thus, the client 
will be the puzzle solver. On the other hand, when $m > \frac{n}{2}$, $\mathbf{C}_m$ 
will be the puzzle solver asking the client to send pieces from the chunks it received 
from honest caches. When $m = \frac{n}{2}$, either party can be the puzzle solver.

Analyzing the bandwidth cost of the above strategy allows us to
configure the number of
puzzle rounds to obtain a specific $\delta$-bound. In order to do so, 
we compute the expected number of
pieces $\mathbb{E}[Y]$ the colluding group will retrieve as 
a function of $R_{puzzle}$ and the number of malicious caches $m$.
Then, we compute $\delta = \frac{\mathbb{E}[Y]}{n \cdot pieces_{total}}$,
after which we select $R_{puzzle}$ that satisfies the desired
$\delta$-bound. To compute $\mathbb{E}[Y]$,
we conduct simulations in which we mimic the
above strategy and track the number of retrieved pieces. As an
example, we consider the following setup,
which we believe is similar to what is used in
practical content distribution applications. We set $chunk_{size} = 1$ MB, and
$piece_{size} = 16$ bytes, leading to $pieces_{total} = 2^{16}$ pieces.
We have $R_{puzzle} \in \{1, \dots, 10\}$,
$n = 6$, and $m \in \{1, \dots, 6\}$. The simulations are repeated
$10^3$ times, where $\mathbb{E}[Y]$, and consequently $\delta$, is computed
as the average across all runs. We also report the standard
deviation of our measurements. The computed $\delta$ values are
found in Table~\ref{simu-results-vary-solver}.

As shown, as the number of rounds increases, $\delta$ increases.
This is expected because a larger number of puzzle rounds means that the challenge
puzzle requires a larger number of pieces to be solved.
Consequently, the puzzle solver is expected to retrieve more content
in order to find these pieces. On the other hand, $\delta$ decreases
as the number of malicious
caches increases for a fixed $R_{puzzle}$ value. Again, this is
expected because more malicious caches makes the collusion
more effective.

\begin{table}[t!]
%\vspace{-24 pt}
\caption{The $\delta$-bound for various $m$ and $R_{puzzle}$ values, $n=6$ caches ($R$ is $R_{puzzle}$). For $m<3$ the client is a more efficient puzzle solver, for $m > 3$ $\mathbf{C}_m$ is a more efficient puzzle solver, $m=3$ is equivalent for each. }
\label{simu-results-vary-solver}
\centering
{\small
\begin{tabular}{| p{0.1\columnwidth}  || p{0.07\columnwidth} | p{0.13\columnwidth} | p{0.13\columnwidth} |  p{0.13\columnwidth} | p{0.13\columnwidth} | p{0.13\columnwidth}| p{0.07\columnwidth}|} \hline
 & \multicolumn{3}{c|}{Client as solver} & Either & \multicolumn{3}{c|}{Cache as solver} \\[0.5ex] \hline
\backslashbox{$R$}{$m$} & {\bf 0} & {\bf 1} & {\bf 2} & {\bf 3} & {\bf 4} & {\bf 5} & {\bf 6} \\ [0.5ex] \hline\hline

{\bf 1} & 1 & 0.87$\mathsmaller{\pm}$0.03 & 0.78$\mathsmaller{\pm}$0.06 & 0.71$\mathsmaller{\pm}$0.08 & 0.45$\mathsmaller{\pm}$0.06 & 0.21$\mathsmaller{\pm}$0.03 & 0 \\[0.5ex] \hline

{\bf 2} & 1 &  0.91$\mathsmaller{\pm}$0.04 & 0.86$\mathsmaller{\pm}$0.06 & 0.82$\mathsmaller{\pm}$0.08 & 0.52$\mathsmaller{\pm}$0.06 &  0.24$\mathsmaller{\pm}$0.04 &   0 \\[0.5ex] \hline

{\bf 3} & 1 & 0.93$\mathsmaller{\pm}$0.04 & 0.9$\mathsmaller{\pm}$0.05 & 0.87$\mathsmaller{\pm}$0.07 & 0.57 $\mathsmaller{\pm}$0.05  &  0.26$\mathsmaller{\pm}$0.04 &  0 \\ [0.5ex]  \hline

{\bf 4} & 1 &  0.94$\mathsmaller{\pm}$0.03 & 0.92$\mathsmaller{\pm}$0.05 & 0.91$\mathsmaller{\pm}$0.06 & 0.59$\mathsmaller{\pm}$0.05 & 0.28$\mathsmaller{\pm}$0.03 & 0 \\ [0.5ex]  \hline

{\bf 5} & 1 &  0.95$\mathsmaller{\pm}$0.03 & 0.94$\mathsmaller{\pm}$0.04 & 0.93$\mathsmaller{\pm}$0.04 & 0.6$\mathsmaller{\pm}$0.05 & 0.29$\mathsmaller{\pm}$0.03 & 0  \\ [0.5ex]  \hline

{\bf 6} & 1 &  0.96$\mathsmaller{\pm}$0.03 & 0.95$\mathsmaller{\pm}$0.04 & 0.94$\mathsmaller{\pm}$0.04 & 0.61$\mathsmaller{\pm}$0.04 &  0.29$\mathsmaller{\pm}$0.03 & 0 \\ [0.5ex]  \hline

{\bf 7} & 1 &  0.96$\mathsmaller{\pm}$0.02 & 0.95$\mathsmaller{\pm}$0.02 & 0.95$\mathsmaller{\pm}$0.04 & 0.62$\mathsmaller{\pm}$0.04 &  0.3$\mathsmaller{\pm}$0.03 &  0 \\ [0.5ex]  \hline

{\bf 8} & 1 &  0.97$\mathsmaller{\pm}$0.02 & 0.96$\mathsmaller{\pm}$0.03 & 0.95$\mathsmaller{\pm}$0.03 & 0.63$\mathsmaller{\pm}$0.03 &  0.3$\mathsmaller{\pm}$0.02 &  0  \\ [0.5ex]  \hline

{\bf 9} & 1 &  0.97$\mathsmaller{\pm}$0.02 & 0.97$\mathsmaller{\pm}$0.03 & 0.96$\mathsmaller{\pm}$0.03 & 0.63$\mathsmaller{\pm}$0.03 &  0.3$\mathsmaller{\pm}$0.02 &  0 \\ [0.5ex]  \hline

{\bf 10} & 1 &  0.97$\mathsmaller{\pm}$0.02 & 0.97$\mathsmaller{\pm}$0.03 & 0.97$\mathsmaller{\pm}$0.03 & 0.64$\mathsmaller{\pm}$0.03 & 0.31$\mathsmaller{\pm}$0.02 & 0 \\ [0.5ex]  \hline

\end{tabular}}
\end{table}

Note that scenarios where $\mathbf{C}_m$ is the solver
have significantly lower $\delta$ values. This is because $\mathbf{C}_m$ 
already possesses the majority of the content that has been pooled 
at no cost (since we assume that metadata retrieval,
such as keys, is free, pooling session keys costs no bandwidth). 
To solve the challenge puzzle, $\mathbf{C}_m$ only needs to retrieve the
missing pieces from the client who has the data chunks from the
honest caches.

Notice that as one would expect, once
the piece provider and solver have their pieces it does not matter which
party was a client or a cache originally.  To see this, examine
$\delta$ in Table~\ref{simu-results-vary-solver}.  Notice that 
the points with $m=2$ and $m=4$ have a difference of $0.33$ (within the margin
of error).  This is because the attacker either retrieves $\frac{2}{6}$ or
$\frac{4}{6}$ of the content, a difference of $0.33$.  However, in either
case, a party with $\frac{4}{6}$ of the content acts as a solver while
the party with $\frac{2}{6}$ provides pieces.  Similar symmetries are found
with $m=1$ and $m=5$ (a difference of $0.66$) and $m=0$ and $m=6$ (a difference of 1). 
This illustrates that once a piece provider and solver have their content,
the cost of solving the puzzle is uniform regardless of which party is the
client and which is a cache.

\section{Evaluation}
\label{eval}
In order to understand how \sysname's security impacts efficiency, this
section evaluates performance in the context of content distribution
applications. Given that \sysname imposes a minimal bandwidth cost to exchange
a puzzle challenge and its solution, what is left to measure is its
computational overhead. Towards this end, we conduct empirical experiments to
answer the following specific questions:
\begin{itemize}
\itemsep0em
\vspace{-2pt}
\item How fast can a publisher generate challenge puzzles?

\item How quickly can a client solve these puzzles?

\item How does the configuration of the design parameters affect these results?

\item What do these numbers mean for a practical deployment?
\end{itemize}

The rest of this section describes our methodology and discusses 
the significance of the obtained results.

\subsection{Methodology}
To establish our benchmarks, we measured the rate, in puzzles per
second, at which a publisher can generate challenge puzzles, and 
the rate at which a client can solve these puzzles.
For the publisher, we considered the
case of popular content that large numbers of clients routinely request in
close time intervals.  For the client, we computed the puzzle solving
rate based on the average case, meaning
that a client tries half the starting pieces in the first data chunk to
find the solution.

Our experiments were conducted on a modest publisher 
server with an AMD Ryzen 3 2200G processor and 16 GiB of
memory, and a low-end client machine with an Intel 
Core i7-4600U processor and 8 GiB of memory.  Each puzzle
generator and solver has been called at least 5,000,000 and 
5,000 times, respectively.  Unless otherwise mentioned, all graphs use 
$R_{puzzle} = 5$, $chunk_{size} = 1$ MB, $h_{size} = 32$ bytes, and
$piece_{size} = 16$ bytes. In addition, instead of reporting the puzzle 
rate for puzzle generator and solver, we compute
the bitrate at which content can be requested using these puzzles.
Despite both the client and publisher operations being embarrassingly 
parallelizable, we run each on a single core to show the 
per-core performance.

\begin{figure*}[t!]
\centering
\subcaptionbox{ Generator speed, $\delta$ effect.\label{gen-bandwidth-delta}}{\includegraphics[width=.48\textwidth,height=1.7in]{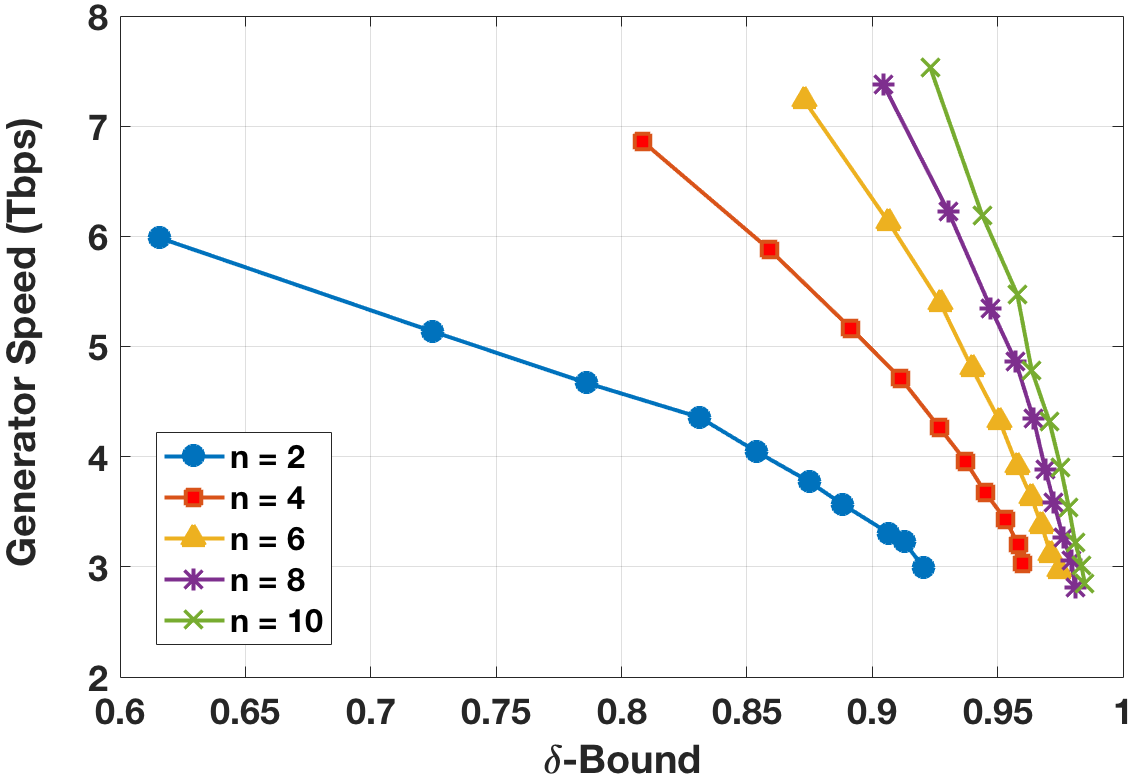}\vspace{-3pt}}\hfill
\subcaptionbox{Generator speed, $\emph{chunk}_{\emph{size}}$ effect\label{gen-bandwidth-blocks}}{\includegraphics[width=.48\textwidth,height=1.7in]{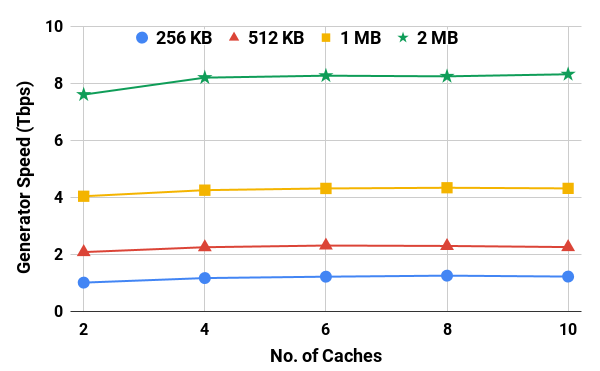}\vspace{-3pt}}\\
\subcaptionbox{Generator speed, $piece_{size}$ effect\label{gen-bandwidth-pieces}}{\includegraphics[width=.48\textwidth,height=1.7in]{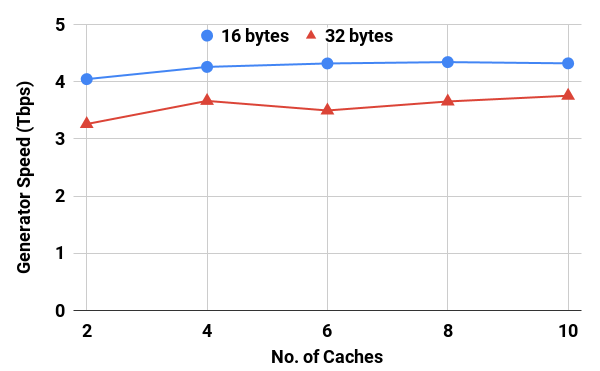}\vspace{-3pt}}\\
\vspace{-6pt}
\caption{Generator speed for various configurations ($n$ is the number of caches in a service session).}
\label{generator-speed}
\vspace{-6pt}
\end{figure*}

\subsection{Results}
\vspace{-8pt}
{\bf Publisher's bitrate vs $\delta$.}  We begin by
measuring the puzzle generation rate   
while varying the number of puzzle rounds $R_{puzzle}$ and number 
of caches $n$ with one malicious cache  
(Figure~\ref{gen-bandwidth-delta}). As shown in the figure, we compute 
the $\delta$-bound value that corresponds to each $R_{puzzle}$ value. 
This produced a curve from 1 round (upper
left point on each curve) to 10 rounds (lower right point). The bitrate 
decreases as $\delta$ increases because with larger $R_{puzzle}$ the publisher 
processes a larger number of pieces when preparing
the challenge, which reduces the puzzle generation rate. On the 
other hand, an increased 
number of caches $n$ increases the throughput
because more data is served per challenge puzzle. This factor also affects the
$\delta$-bound of \sysname. As shown in the figure, for larger $n$
the range $\delta$ gets larger for all $R_{puzzle}$ values. That is, the
impact of having a malicious cache decreases when $n$ gets larger. This 
captures what happens in 
real life, where it will be harder for caches to collude effectively when there is a
large number of caches per session.
Based on the figure, setting $R_{puzzle} \geq 5$ in practice provides a reasonable
$\delta$-bound for $n \geq 4$ ($\delta \geq 0.93$), with diminishing 
returns thereafter.\\

\begin{figure*}[t!]
\centering
\subcaptionbox{Solver speed, $\delta$ effect.\label{solver-bandwidth-delta}}{\includegraphics[width=.48\textwidth,height=1.7in]{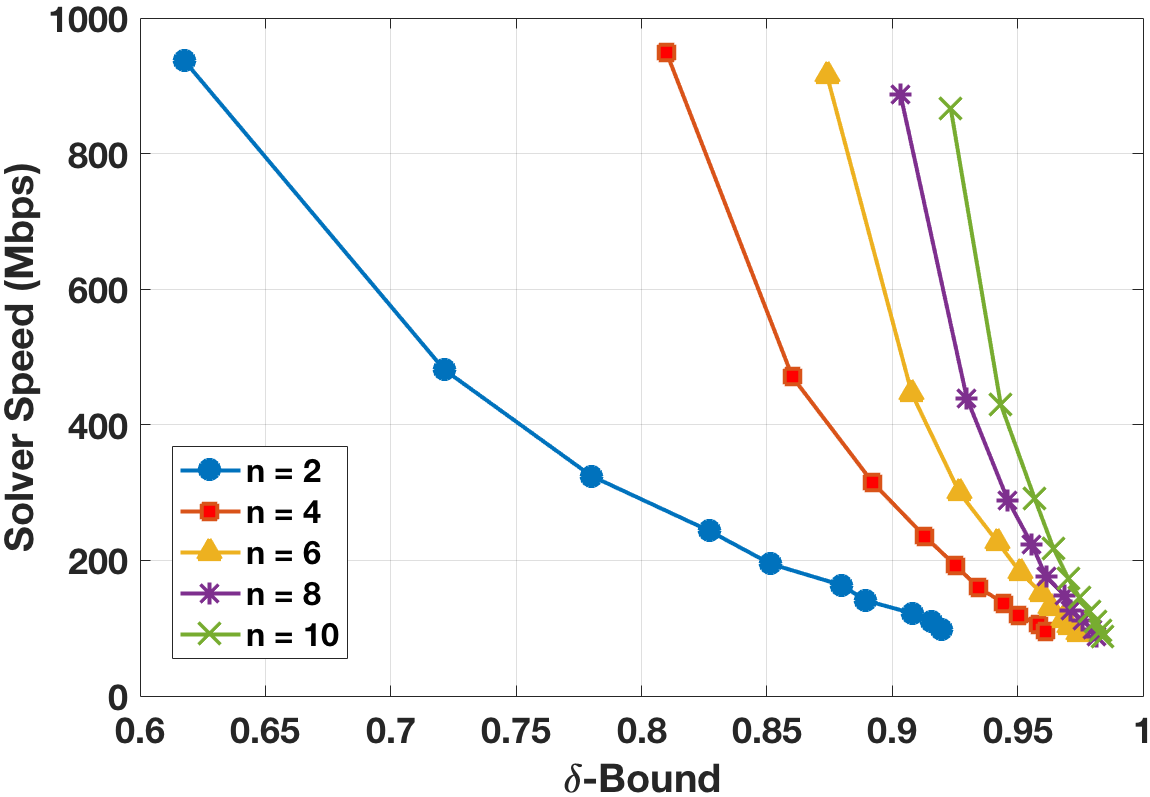}\vspace{-3pt}}\hfill
\subcaptionbox{Solver speed, $chunk_{size}$ effect\label{solver-bandwidth-blocks}}{\includegraphics[width=.48\textwidth,height=1.7in]{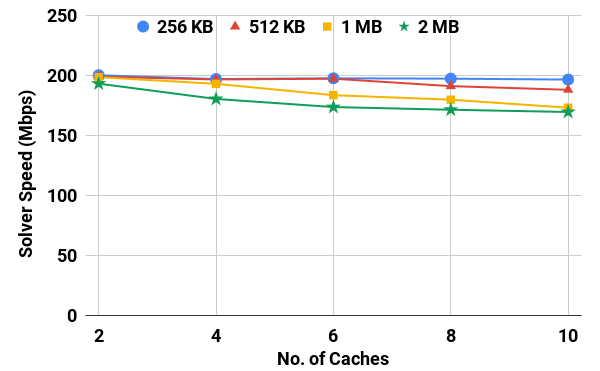}\vspace{-3pt}}\\
\subcaptionbox{Solver speed, $piece_{size}$ effect\label{solver-bandwidth-pieces}}{\includegraphics[width=.48\textwidth,height=1.7in]{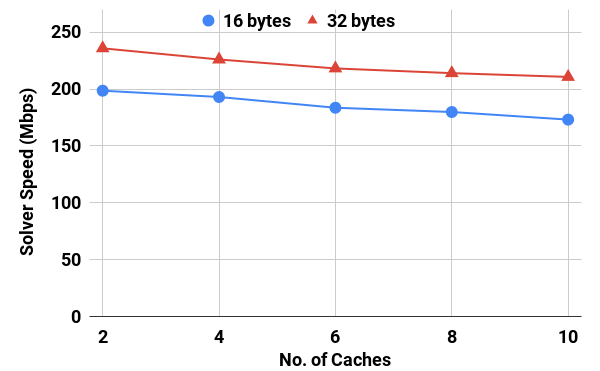}\vspace{-3pt}}\\
\caption{Solver speed for various configurations ($n$ is the number of caches in a service session).}
\label{solver-speed}
\end{figure*}

{\bf Client's bitrate vs $\delta$.} Figure~\ref{solver-bandwidth-delta} 
shows the bitrate vs $\delta$-bound for the puzzle solver. As shown,   
for a fixed $R_{puzzle}$ value, the client's effective bandwidth is 
relatively uniform independent of the number of caches $n$.
However, the $R_{puzzle}$ value, for a fixed $n$, has substantial impact on the
effective bandwidth of a client. Given that the reported speed in the figure 
dwarfs the 5 Mbps Netflix 1080p quality video rate~\cite{netflix-speed}, 
even using $R_{puzzle} = 5$ (the 5$^{th}$ point on each curve starting from the left), 
our modest client machine is able to watch dozens of 1080p videos concurrently.
If higher performance is desired, then reducing $R_{puzzle}$, i.e., reducing $\delta$,
provides drastically better performance, up to 900Mbps, if needed.\\

{\bf How does $chunk_{size}$ impact client and publisher bitrates?}
We measured the puzzle generation and solving rates for
various data chunk sizes and $n$ values (results are shown in
Figures \ref{gen-bandwidth-blocks} and \ref{solver-bandwidth-blocks}).
The $chunk_{size}$ has a large effect on the performance of publishers, but
minimal effect on client's performance. There are several reasons for this 
difference.
The publisher has almost a fixed puzzle generation
rate regardless of the chunk size, because it 
processes the same number of pieces for fixed $R_{puzzle}$, $piece_{size}$,
and $n$ values. 
Consequently, a larger $chunk_{size}$ makes the amount of content
served per challenge puzzle larger. Alternatively, for a client the puzzle solving
rate decreases with larger $chunk_{size}$ because the client has to try a
larger number of trial puzzles. When computing the bitrate for some $n$ value,
the low puzzle rates are multiplied by large $chunk_{size}$ and vice versa. For this reason
the client bandwidth is somewhat similar for all $chunk_{size}$ values.\\

{\bf How does $piece_{size}$ impact client and publisher bitrates?}
To understand how to set the piece size, we studied its effect
on publisher (Figure~\ref{gen-bandwidth-pieces}) and client
(Figure~\ref{solver-bandwidth-pieces}) performance.  The publisher can generate
enough puzzles to serve over 3 Tbps, regardless of the piece size. However,
a piece size of 16 bytes is slightly more efficient because AES-CTR works on
16 byte blocks for encryption. In addition, a smaller piece size means that the
publisher processes a smaller amount of content for a fixed number of
pieces. The client, on the other hand, tends to benefit
from larger piece sizes because they reduce the number of starting pieces, 
and hence, trial puzzles, a client has to compute.  Given that
a client with $piece_{size} = 16$ byte already has a high throughput,
and given that publishers are usually heavy-loaded entities,
we recommend the use of $piece_{size} = 16$ bytes to boost 
publisher performance.\\

{\bf Contextualizing our results.} To ground our results in real world
numbers, we examined the customer demand from the popular content 
provider Netflix.com. Netflix serves 1080p video at a bitrate of
approximately 5 Mbps~\cite{netflix-speed}.  As shown in
Figure \ref{gen-bandwidth-blocks}, and taking $n = 4$ caches, a 
publisher in our setup, using a single core machine, can generate puzzles 
for 136,000 requests per second, which translates to 544,000 data 
chunks per second.  
To understand this load, we look at a popular show, ``House of Cards'', 
where 5.4 million of its 83 million subscribers (as of 2015) watched at least 
one episode within a month of its release~\cite{houseofcards}.  Since the
report does not indicate how many of those views were concurrent, it is 
not possible to infer the exact peak load.  However,
our single core publisher supports concurrent viewing from 870,000 clients
which is enough to support a simultaneous viewing peak from 1/6 of the
total views at any point during the first month.

As for the client, the previous results showed that for $R_{puzzle} = 5$, 
on average our 
low-end client is able to solve enough puzzles to retrieve at least 170 
Mbps using a single core (Figure~\ref{solver-bandwidth-blocks}).  
This is more than 34 times the rate required to
retrieve the same popular 1080p video~\cite{netflix-speed}.\\

In summary, the previous results demonstrate that \sysname is
computationally-lightweight. Its security in fighting cache accounting
attacks is substantial ($\delta > 0.95$ with generous attacker assumptions),
even at bandwidth values that support a publisher serving thousands of 
clients or a client simultaneously watching dozens of 1080p videos.

\section{Conclusions}
\label{conclusions}
In this paper, we introduce \sysname, a low-overhead solution to 
defend against cache accounting attacks in peer-assisted CDNs. 
\sysname is the first system that forces malicious caches, even when colluding
with clients, to expend substantial bandwidth to demonstrate that content was
retrieved. This is done by introducing a cache accountability puzzle 
that provides strong protections even given unrealistically
strong assumptions about the attacker's capabilities.  For example, 
with a 5 round puzzle, if 3 malicious caches out of 6 total caches wish 
to perform a cache accounting attack, the 
colluding parties would retrieve on average more than 0.95 of the requested 
content ($\delta>0.95$).
We analyze the security of \sysname, and show experimentally 
that it incurs a low computation cost, enabling a modest client to retrieve 
170Mbps from a modest publisher 
serving several Tbps. This demonstrates the viability of 
employing our scheme in large scale content distribution 
applications.

{\footnotesize \bibliographystyle{acm}
\bibliography{puzzleBib}}

%\theendnotes

\end{document}